\documentclass[fleqn,iopfts,12pt]{iopart}
\usepackage{graphicx}
\usepackage{iopams}
\eqnobysec
\begin{document}
\title{Coherent states of a charged particle in a uniform magnetic
field}
\author{K Kowalski and J Rembieli\'nski}
\address{Department of Theoretical Physics, University
of \L\'od\'z, ul.\ Pomorska 149/153, 90-236 \L\'od\'z,
Poland}
\begin{abstract}
The coherent states are constructed for a charged particle in a
uniform magnetic field based on coherent states for the circular
motion which have recently been introduced by the authors.
\end{abstract}
\submitto{\JPA}
\maketitle
\section{Introduction}
Coherent states which can be regarded from the physical point of
view as the states closest to the classical ones, are of fundamental
importance in quantum physics.  One of the most extensively studied
quantum systems presented in many textbooks is a charged particle in
a uniform magnetic field.  The coherent states for this system were
originally found by Malkin and Man'ko \cite{1} (see also Feldman and
Kahn \cite{2}).  As a matter of fact, the alternative states for a
charged particle in a constant magnetic field were introduced by
Loyola, Moshinsky and Szczepaniak \cite{3} (see also the very recent
paper by Schuch and Moshinsky \cite{4}), nevertheless, those states
are labeled by discrete quantum numbers and therefore can hardly be
called ``coherent ones'' which should be marked with the points of
the classical phase space.  In spite of the fact that the transverse
motion of a charged particle in in a uniform magnetic field is
circular, the coherent states described by Malkin and Man'ko are
related to the standard coherent states for a particle on a plane
instead of coherent states for a particle on a circle.  Furthermore,
the definition of the coherent states constructed by Malkin and Man'ko
seems to ignore the momentum part of the classical phase space.  In
this work we introduce the coherent states for a charged particle in
a uniform magnetic field based on the construction of the coherent
states for a quantum particle on a circle described in \cite{5}.  The
paper is organized as follows.  In section 2 we recall the construction 
of the coherent states for a particle on a circle.  Section 3
summarizes the main facts about quantization of a charged particle
in a magnetic field.  Section 4 is devoted to the definition of the
coherent states for a charged particle in a magnetic field and
discussion of their most important properties.  In section 5 we
collect the basic facts about the coherent states for a charged particle
in a magnetic field introduced by Malkin and Man'ko and we compare
these states with ours discussed in section 4.
\section{Coherent states for a quantum mechanics on a circle}
In this section we summarize most important facts about the coherent
states for a quantum particle on a circle.  We first recall that the
algebra adequate for the study of the motion on a circle is of the
form 
\begin{equation}
[J,U] = U,\qquad [J,U^\dagger]=-U^\dagger ,
\end{equation}
where $J$ is the angular momentum operator, the unitary operator
$U$ represents the position of a quantum particle on a (unit) circle
and we set $\hbar=1$.  Consider the eigenvalue equation
\begin{equation}
J|j\rangle = j|j\rangle.
\end{equation}
As shown in \cite{5} $j$ can be only integer and half-integer.  We
restrict for brevity to the case with integer $j$.  From (2.1) and
(2.2) it follows that the operators $U$ and $U^\dagger $ are the 
ladder operators, namely
\begin{equation}
U|j\rangle = |j+1\rangle,\qquad U^\dagger |j\rangle = |j-1\rangle.
\end{equation}
Consider now the coherent states for a quantum particle on a circle.
These states can be defined \cite{5} as the solution of the eigenvalue equation
\begin{equation}
X|\xi\rangle = \xi|\xi\rangle,
\end{equation}
where 
\begin{equation}
X = e^{-J +\frac{1}{2}}U.
\end{equation}
An alternative construction of the coherent states specified by (2.4)
based on the Weil-Brezin-Zak transform was described in \cite{6}.
The convenient parametrization of the complex number $\xi$ consistent 
with the form of the operator $X$ is given by
\begin{equation}
\xi = e^{-l + \rmi\varphi}.
\end{equation}
The parametrization (2.6) arises from the deformation of the
cylinder (the phase space) specified by
\begin{equation}
x=e^{-l}\cos\varphi,\qquad y=e^{-l}\sin\varphi,\qquad z=l,
\end{equation}
and then projecting the points of the obtained surface on the $x,y$ 
plane.  The projection of the vectors $|\xi\rangle$ onto the basis vectors
$ |j\rangle$ is of the form
\begin{equation}
\langle j|\xi\rangle = \xi^{-j}e^{-\frac{j^2}{2}}.
\end{equation}
Using the parameters $l$, and $\varphi$ (2.8) can written in the following 
equivalent form:
\begin{equation}
\langle j|l,\varphi\rangle =
e^{lj-\rmi j\varphi}e^{-\frac{j^2}{2}},
\end{equation}
where $|l,\varphi\rangle\equiv|\xi\rangle$, with $\xi=e^{-l + \rmi\varphi}$.
The coherent states are not orthogonal.  Namely,
\begin{equation}
\langle \xi|\eta\rangle =
\sum_{j=-\infty}^{\infty}(\xi^*\eta)^{-j}e^{-j^2} =
\theta_3\left(\frac{\rmi}{2\pi}\ln\xi^*\eta\Bigg\vert\frac{\rmi}
{\pi}\right),
\end{equation}
where $\theta_3$ is the Jacobi theta-function \cite{7}.  The coherent states 
satisfy
\begin{equation}
\frac{\langle l,\varphi|J|l,\varphi\rangle}{\langle
l,\varphi|l,\varphi\rangle}\approx l,
\end{equation}
where the maximal error is of order $0.1$ per cent and we have the exact 
equality in the case with $l$ integer or half-integer.  Therefore, the 
parameter $l$ labeling the coherent states can be interpreted as the
classical angular momentum.  Furthermore, we have
\begin{equation}
\frac{\langle l,\varphi|U|l,\varphi\rangle}{\langle l,\varphi|l,\varphi\rangle}
 \approx
e^{-\frac{1}{4}}e^{\rmi\varphi}.
\end{equation}
We point out that the absolute value of the average of the unitary operator 
$U$ given by (2.12) which is approximately $e^{-\frac{1}{4}}$ is lesser than
1, as expected because $U$ is not diagonal in the coherent states basis.  On 
introducing the relative expectation value
\begin{equation}
\langle\!\langle U\rangle\!\rangle_{(l,\varphi)} :=
\frac{\langle U\rangle_{(l,\varphi)}}{\langle U\rangle_{(0,0)}},
\end{equation}
where $\langle U\rangle_{(l,\varphi)}=\langle
l,\varphi|U|l,\varphi\rangle/\langle l,\varphi|l,\varphi\rangle$, we
get
\begin{equation}
\langle\!\langle U\rangle\!\rangle_{(l,\varphi)} \approx e^{\rmi\varphi}.
\end{equation}
Therefore, the relative expectation value $\langle\!\langle
U\rangle\!\rangle_{(l,\varphi)}$ seems to be the most natural
candidate to describe the average position on a circle and $\varphi$
can be regarded as the classical angle.  We finally point out that
the discussed coherent states as well as the coherent states for a
particle on a sphere introduced by us in \cite{8} are concrete
realization of the general mathematical scheme of construction of
the Bargmann spaces described in the recent papers \cite{9}.  The
importance of the coherent states for the circular motion has been
confirmed by their recent application in quantum gravity \cite{10}.
\section{Charged quantum particle in a magnetic field}
In order to obtain the operators necessary for definition of the
coherent states we first recall some facts about the quantization of
a particle with the mass $\mu$ and a charge $e$ in a uniform magnetic
field ${\bi B}=(0,0,B)$, which is taken, without loss of generality,
along the $z$ axis.  Neglecting the spin we can write the
Hamiltonian in the form
\begin{equation}
H = \frac{1}{2\mu}{\bpi}^2,
\end{equation}
where $\bpi=\mu\dot{\bi x}$ is the kinetic momentum related to
the canonical momentum ${\bi p}$ satisfying the Heisenberg algebra with
the position ${\bi x}$, by
\begin{equation}
\bpi = {\bi p} - e {\bi A},
\end{equation}
where ${\bi A}$ is the vector potential which fulfils ${\bi B}=\hbox{rot}
{\bi A}$ and we set $c=1$.  We choose the symmetric gauge such that
\begin{equation}
{\bi A} = (-By/2,Bx/2,0)
\end{equation}
in which ${\bi A}=\frac{1}{2}{\bi B}\times{\bi x}$.  The coordinates
of the kinetic momentum (3.2) in the gauge (3.3) are
\begin{equation}
\pi_x = p_x+\frac{\mu\omega}{2}y,\qquad \pi_y = p_y-\frac{\mu\omega}{2}x,
\qquad \pi_z=p_z,
\end{equation}
where $\omega =\frac{eB}{\mu}$ is the cyclotron frequency.  From (3.1)
and(3.4) it follows that the motion along the $z$ axis is free and
we actually deal with a two-dimensional problem in the $x,y$
plane.  Clearly, the Hamiltonian for the transverse motion is
\begin{equation}
H_{\perp} = \frac{1}{2\mu}(\pi_x^2+\pi_y^2).
\end{equation}
The coordinates $\pi_x$ and $\pi_y$ of the kinetic momentum given by
(3.4) satisfy the following commutation relation:
\begin{equation}
[\pi_x,\pi_y] = \rmi\mu\omega ,
\end{equation}
where we set $\hbar=1$.  On introducing the operators
\begin{equation}
a = \frac{1}{\sqrt{2\mu\omega}}(-\pi_y+\rmi\pi_x),\qquad
a^\dagger = \frac{1}{\sqrt{2\mu\omega}}(-\pi_y-\rmi\pi_x),
\end{equation}
which obey
\begin{equation}
[a,a^\dagger]=1,
\end{equation}
we can write the Hamiltonian (3.5) in the form of the Hamiltonian of the 
harmonic oscillator, such that
\begin{equation}
H_{\perp} = \omega (a^\dagger a + \hbox{$\scriptstyle 1\over 2$}).
\end{equation}
Consider now the orbit center-coordinate operators \cite{11}
\begin{equation}
x_0 = x + \frac{1}{\mu\omega}\pi_y,\qquad y_0 = y -
\frac{1}{\mu\omega}\pi_x.
\end{equation}
These operators are integrals of the motion and they represent the
coordinates of the center of a circle in the $x,y$ plane in
which a particle moves.  However, they do not commute with each
other, namely, we have
\begin{equation}
[x_0,y_0] = -\frac{\rmi}{\mu\omega}.
\end{equation}
As with coordinates of the kinetic momentum we can construct from
$x_0$ and $y_0$ the creation and annihilation operators.  We set
\begin{equation}
b=\sqrt{\frac{\mu\omega}{2}}(x_0-\rmi y_0),\qquad
b^\dagger =\sqrt{\frac{\mu\omega}{2}}(x_0+\rmi y_0),
\end{equation}
implying with the use of (3.11)
\begin{equation}
[b,b^\dagger] = 1.
\end{equation}
We now return to (3.10).  Since equations (3.10) hold also in the
classical case, therefore the operators
\begin{equation}
r_x := x-x_0=-\frac{1}{\mu\omega}\pi_y,\qquad
r_y := y-y_0=\frac{1}{\mu\omega}\pi_x,
\end{equation}
are the position observables of a particle on a circle.  More
precisely, they are coordinates of the radius vector of a particle
moving in a circle with the center at the point $(x_0,y_0)$.  From 
(3.14) and (3.6) it follows that
\begin{equation}
[r_x,r_y] = \frac{\rmi}{\mu\omega}.
\end{equation}
We have the formula on the squared radius of a circle such that
\begin{equation}
{\bi
r}^2=r_x^2+r_y^2=(x-x_0)^2+(y-y_0)^2=\frac{1}{(\mu\omega)^2}
(\pi_x^2+\pi_y^2)=\frac{2}{\mu\omega^2}H_{\perp}
\end{equation}
following directly from (3.10) and (3.5).
\section{Coherent states for a particle in a magnetic field}
An experience with the coherent states for a circular motion
described in section 2 indicates that in order to introduce the
coherent states we should first identify the algebra adequate for
the study of the motion of a charged particle in a uniform magnetic
field.  As with (2.1) such algebra should include the angular
momentum operator.  It seems that the most natural candidate is the
operator defined by
\begin{equation}
L = ({\bi r}\times\bpi)_z = r_x\pi_y-r_y\pi_x.
\end{equation}
Indeed, eqs.\ (3.14) and (3.16) taken together yield
\begin{equation}
L = -\mu\omega {\bi r}^2,
\end{equation}
which coincides with the classical expression.  Furthermore, it can
be easily verified that it commutes with the orbit center-coordinate
operators $x_0$ and $y_0$.  It should be noted however that since
\begin{equation}
[L,r_x] = 2\rmi r_y,\qquad [L,r_y] = -2\rmi r_x,
\end{equation}
following directly from (4.2) and (3.15), the generator of rotations
about the axis passing through the center of the circle and
perpendicular to the $x,y$ plane, is not $L$ but $\frac{1}{2}L$.
Therefore, the counterpart of the operator $J$ satisfying (2.1)
which is the generator of the rotations, is not $L$ but $\frac{1}{2}L$.

Now, we introduce the operator representing the position of a
particle on a circle of the form
\begin{equation}
r_+ = r_x+\rmi r_y.
\end{equation}
This operator is a natural counterpart of the unitary operator $U$
representing the position of a quantum particle on a unit circle
discussed in section 2.  Clearly, the algebra should include the
orbit-center operators $x_0$ and $y_0$.  Bearing in mind the
parametrization (4.4) it is plausible to introduce the operator
\begin{equation}
r_{0+} = x_0+\rmi y_0
\end{equation}
which has the meaning of the operator corresponding to the center
of the circle.  In order to complete the algebra we also introduce the
Hermitian conjugates of the operators $r_+$ and $r_{0+}$,
respectively, such that
\begin{equation}
r_- = r_x-\rmi r_y,\qquad r_{0-} = x_0-\rmi y_0.
\end{equation}
Taking into account (4.3), (3.15) and (3.11) we arrive at the
following algebra which seems to be most natural in the case of the
circular motion of a charged particle in a uniform magnetic field:
\begin{equation}
\fl [L,r_\pm]=\pm 2r_\pm,\quad [L,r_{0\pm}]=0,\quad
[r_+,r_-]=\frac{2}{\mu\omega},\quad [r_{0+},r_{0-}]=-\frac{2}{\mu\omega},
\quad [r_{\pm},r_{0\pm}]=0.
\end{equation}
The algebra (4.7) has the Casimir operator given in the unitary
irreducible representation by
\begin{equation}
r_-r_+ + \frac{1}{\mu\omega}L = cI,
\end{equation}
where $c$ is a constant.  We choose the representation referring to
$c=-\frac{1}{\mu\omega}$ because it is the only one such that (4.8)
with $r_\pm$ given by (4.4) and (4.6) is equivalent to (4.2).
Consider now the creation and annihilation operators defined by
\begin{equation}
a=\sqrt{\frac{\mu\omega}{2}}r_+,\qquad
a^\dagger=\sqrt{\frac{\mu\omega}{2}}r_-,
\end{equation}
which coincide in view of (4.4) and (3.14) with the operators
(3.7).  The Casimir (4.8) with $c=-\frac{1}{m\omega}$ written with
the help of the Bose operators (4.9) takes the form
\begin{equation}
L = -(2N_a+1),
\end{equation}
where $N_a=a^\dagger a$.  Furthermore, it follows from (4.7) that
the creation and annihilation operators such that (see (3.12), (4.5)
and (4.6))
\begin{equation}
b = \sqrt{\frac{\mu\omega}{2}}r_{0-},\qquad b^\dagger = 
\sqrt{\frac{\mu\omega}{2}}r_{0+}
\end{equation}
commute with $a$ and $a^\dagger$.  Therefore, the operators
$N_a=a^\dagger a$ and $N_b=b^\dagger b$, commute with each other.
Consider the irreducible representation of the algebra (4.7) spanned
by the common eigenvectors of the number operators $N_a$ and $N_b$
satisfying
\begin{equation}
N_a |n,m\rangle = n|n,m\rangle,\qquad N_b |n,m\rangle = m|n,m\rangle,
\end{equation}
where $n$ and $m$ are nonnegative integers.  Using (4.10), (4.9) and
(4.11) we find that the generators of the algebra (4.7) act on the
basis vectors $|n,m\rangle$ in the following way:
\numparts
\begin{eqnarray}
L|n,m\rangle &=& -(2n+1)|n,m\rangle,\\
r_+|n,m\rangle &=& \sqrt{\frac{2n}{\mu\omega}}|n-1,m\rangle,\\
r_-|n,m\rangle &=& \sqrt{\frac{2(n+1)}{\mu\omega}}|n+1,m\rangle,\\
r_{0+}|n,m\rangle &=& \sqrt{\frac{2(m+1)}{\mu\omega}}|n,m+1\rangle,\\
r_{0-}|n,m\rangle &=& \sqrt{\frac{2m}{\mu\omega}}|n,m-1\rangle.
\end{eqnarray}
\endnumparts
Now bearing in mind the form of the eigenvalue equation (2.4) and
the discussion above we define the coherent states for a charged
particle in a uniform magnetic field as the simultaneous
eigenvectors of the commuting non-Hermitian operators $Z$ and
$r_{0-}$:
\numparts
\begin{eqnarray}
&&Z|\zeta,z_0\rangle = \zeta |\zeta,z_0\rangle,\\
&&r_{0-}|\zeta,z_0\rangle = z_0|\zeta,z_0\rangle,
\end{eqnarray}
\endnumparts
where
\begin{equation}
Z=e^{-\frac{L}{2}+\frac{1}{2}}r_+,
\end{equation}
and we recall that $r_{0-}$ is proportional to the Bose annihilation
operator $b$ (see (4.11)), so that the coherent states
$|\zeta,z_0\rangle$ can be viewed as tensor product of the
eigenvectors $|\zeta\rangle$ of the operator $Z$ and the standard
coherent states $|z_0\rangle$.  Clearly, the complex number $\zeta$
parametrizes the classical phase space for the circular motion of a
charged particle while the complex number $z_0$ represents the
position of the center of the circle.  Taking into account (4.14) and
(4.13) we find
\begin{equation}
\langle n,m|\zeta,z_0\rangle=\left(\frac{\mu\omega}{2}\right)^\frac{n}{2}
\frac{\zeta^n}{\sqrt{n!}}e^{-\frac{1}{2}(n+\frac{1}{2})^2}
\left(\frac{\mu\omega}{2}\right)^\frac{m}{2}\frac{z_0^m}{\sqrt{m!}}.
\end{equation}
Now, the form of the operator $Z$ and (2.6) indicate the following
parametrization of the complex number $\zeta$:
\begin{equation}
\zeta = r(l)e^{-\frac{l}{2}+\rmi\varphi},
\end{equation}
where $l$ is real non-positive and
$r(l)=\sqrt{-\frac{l}{\mu\omega}}$ is the classical radius of the
circle in which moves a particle implied by the classical relation
$l=-\mu\omega r^2$.  Further, in accordance with (4.6) we set
\begin{equation}
z_0 = \overline x_0 - \rmi\overline y_0,
\end{equation}
where $\overline x_0$ and $\overline y_0$ are real.  Using (4.17)
and (4.18) we can write (4.16) in the form
\begin{equation}
\fl \langle n,m|l,\varphi;\overline x_0,\overline y_0 \rangle =
\left(-\frac{l}{2}e^{-l}\right)^\frac{n}{2}\frac{e^{\rmi n\varphi}}
{\sqrt{n!}}e^{-\frac{1}{2}(n+\frac{1}{2})^2}\frac{1}{\sqrt{m!}}
\left[\frac{1}{\sqrt{2}}\left(\frac{\overline x_0}{\lambda}-\rmi
\frac{\overline y_0}{\lambda}\right)\right]^m,
\end{equation}
where $|l,\varphi;\overline x_0,\overline y_0\rangle\equiv|\zeta,z_0
\rangle$ with $\zeta$ and $z_0$ given by (4.17) and (4.18),
respectively, and $\lambda=1/\sqrt{\mu\omega}$ is the classical
radius of the ground state Landau orbit.

As with the states $|\xi\rangle$ given by (2.4) our most important
criterion to test the correctness of the introduced coherent states 
$|\zeta,z_0\rangle$ will be their closeness to the classical phase
space.  Consider the expectation value of the angular momentum
operator $L$.  Taking into account the completeness of the states 
$|n,m\rangle$, (4.13a) and (4.19) we get
\begin{equation}
\langle L\rangle_l=\frac{\langle l,\varphi;\overline x_0,\overline y_0|L|
l,\varphi;\overline x_0,\overline y_0\rangle}
{\langle l,\varphi;\overline x_0,\overline y_0|
l,\varphi;\overline x_0,\overline y_0\rangle}=
-\frac{\sum_{n=0}^\infty  \frac{2n+1}{n!}\left(-\frac{l}{2}e^{-l}\right)^n
e^{-(n+\frac{1}{2})^2}}
{\sum_{n=0}^\infty  \frac{1}{n!}\left(-\frac{l}{2}e^{-l}\right)^n
e^{-(n+\frac{1}{2})^2}}.
\end{equation}
From computer calculations it follows that $\langle
L\rangle_l\approx l$.  Nevertheless, in opposition to the case of
the coherent states for a quantum particle on a circle discussed in
section 2 the approximate equality of $\langle L\rangle_l$ and $l$
does not hold for practically arbitrary small $|l|$.  More precisely,
we have found that the approximation is very good for $|l|\ge1$ (the
bigger $l$ the better approximation).  For example if $|l|\sim1$
then the relative error $|(\langle L\rangle_l-l)/l|\sim1\%$.  In our
opinion such behavior of $\langle L \rangle_l$ means that for small
$|l|$ the quantum fluctuations are not negligible and the
description based on the concept of the classical phase space is not
an adequate one.  We remark that the same phenomenon have been
observed in the case of the coherent states for a particle on a
sphere \cite{8}.  Thus, it turns out that the parameter $l$ in (4.17)
can be identified (in general approximately) with the classical
angular momentum of a charged particle in a uniform magnetic field.

We now discuss the position of a particle on a circle in the context
of the introduced coherent states.  Using (4.13b) and (4.19) we find
\begin{equation}
\fl \langle r_+\rangle_{(l,\varphi)}=
\frac{\langle l,\varphi;\overline x_0,\overline y_0|r_+|
l,\varphi;\overline x_0,\overline y_0\rangle}
{\langle l,\varphi;\overline x_0,\overline y_0|
l,\varphi;\overline x_0,\overline y_0\rangle}=
r(l)e^{\rmi\varphi}e^{-\frac{1}{4}}e^{-\frac{l}{2}}
\frac{\sum_{n=0}^\infty  \frac{1}{n!}\left(-\frac{l}{2}e^{-l}\right)^n
e^{-(n+1)^2}}
{\sum_{n=0}^\infty  \frac{1}{n!}\left(-\frac{l}{2}e^{-l}\right)^n
e^{-(n+\frac{1}{2})^2}},
\end{equation}
where $r(l)=\sqrt{-\frac{l}{\mu\omega}}$ is the classical formula on
the radius of the circle in which moves a particle (see (4.17)).
The computer calculations indicate that
\begin{equation}
\langle r_+\rangle_{(l,\varphi)}\approx
r(l)e^{\rmi\varphi}e^{-\frac{1}{4}},
\end{equation}
where the approximation is very good but a bit worse than that in
the case with $\langle L\rangle_l$.  Namely, for $|l|=5$ the
relative error is of order $1\%$.  Because of the term
$e^{-\frac{1}{4}}$ it turns out that the average value of $r_+$ does
not belong to the circle with radius $r(l)$.  Motivated by the
formal resemblance of (4.22) with $r=1$ and (2.12) we identify the
correct expectation value as
\begin{equation}
\langle\!\langle r_+ \rangle\!\rangle_{(l,\varphi)}=e^{\frac{1}{4}}
\langle r_+\rangle_{(l,\varphi)}=r(l)e^{\rmi\varphi}e^{-\frac{l}{2}}
\frac{\sum_{n=0}^\infty  \frac{1}{n!}\left(-\frac{l}{2}e^{-l}\right)^n
e^{-(n+1)^2}}
{\sum_{n=0}^\infty  \frac{1}{n!}\left(-\frac{l}{2}e^{-l}\right)^n
e^{-(n+\frac{1}{2})^2}},
\end{equation}
so
\begin{equation}
\langle\!\langle r_+ \rangle\!\rangle_{(l,\varphi)}\approx
r(l)e^{\rmi\varphi}
\end{equation}
which is a counterpart of (2.14).  In our opinion, the appearance of
the same factor $e^{-\frac{1}{4}}$ in formulas (2.12) and (4.22)
confirms the correctness of the approach taken up in this work.  In 
view of the form of (4.24) it appears that $r(l)e^{\rmi\varphi}$ (see 4.17) 
can be interpreted as the classical parametrization of a position of a 
charged particle in a uniform magnetic field.

We now study the distribution of vectors $|n,m\rangle$ in the
normalized coherent state.  The computer calculations indicate that
the function
\begin{equation}
\fl p_{n,m}(l,\overline x_0,\overline y_0) = 
\frac{|\langle n,m|l,\varphi;\overline x_0,\overline y_0 \rangle|^2}
{\langle l,\varphi;\overline x_0,\overline y_0|
l,\varphi;\overline x_0,\overline y_0\rangle}
=\frac{\frac{1}{n!}\left(-\frac{l}{2}e^{-l}\right)^ne^{-(n+\frac{1}{2})^2}
\frac{1}{m!}\left(\frac{\mu\omega}{2}\right)^m(\overline x_0^2+
\overline y_0^2)^m}{\left(\sum_{n=0}^{\infty}\frac{1}{n!}\left(-\frac{l}{2}
e^{-l}\right)^n e^{-(n+\frac{1}{2})^2}\right)e^{\frac{\mu\omega}{2}
(\overline x_0^2+\overline y_0^2)}}
\end{equation}
which gives the probability of finding the system in the state
$|n,m\rangle$ when the system is in normalized coherent state 
$|l,\varphi;\overline x_0,\overline y_0\rangle/\sqrt{
\langle l,\varphi;\overline x_0,\overline y_0|l,\varphi;\overline
x_0,\overline y_0\rangle}$, is peaked for fixed $l$, $m$, $x_0$ and
$y_0$ at point $n_{{\rm max}}$ coinciding with the integer nearest
to $-(l+1)/2$ (see figure 1).  In view of the relation (4.10) this
observation confirms once more the interpretation of the parameter
$l$ as the classical angular momentum.
\begin{figure*}
\centering
\includegraphics[scale=.8]{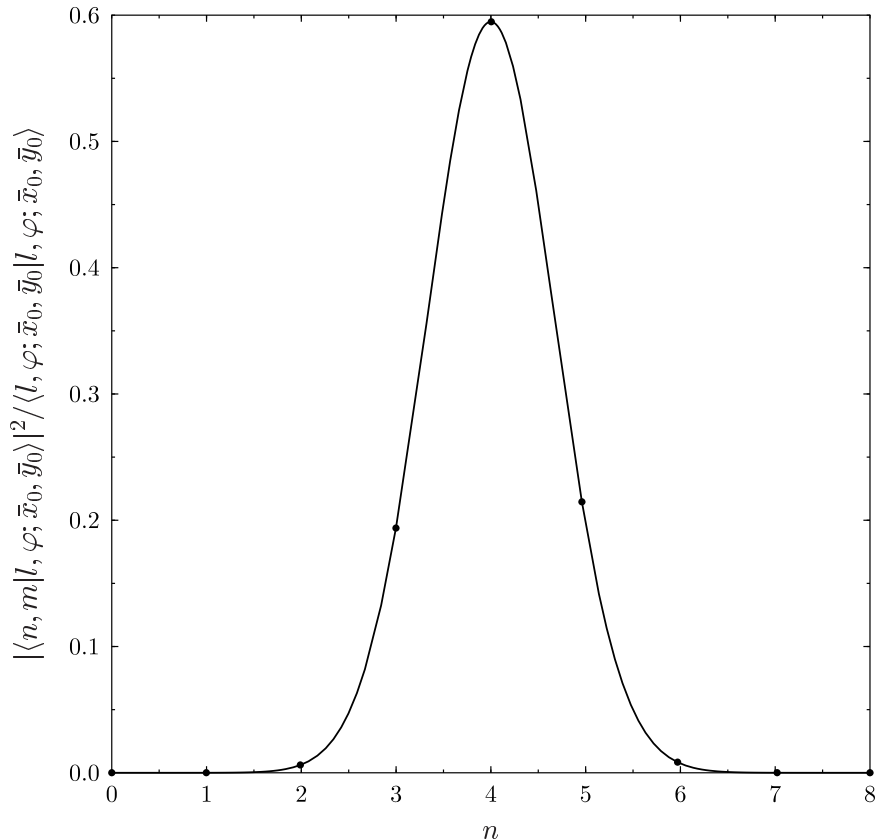}
\caption{The plot of $p_{n,m}(l,\bar x_0,\bar y_0)$ versus $n$ (see
4.25), where $l=-9$, $m=0$, and $\bar x_0~=~~\bar y_0~=~~0$.  The 
maximum is reached at point $n_{\rm max}=4$ coinciding with $-(l+1)/2$.}
\end{figure*}
For the sake of completeness we now write down the formula on the
expectation value of the operator $r_{0-}$ representing the position
of the center of the circle, such that
\begin{equation}
\langle l,\varphi;\overline x_0,\overline y_0|r_{0-}|
l,\varphi;\overline x_0,\overline y_0\rangle = \overline x_0
-\rmi\overline y_0
\end{equation}
following immediately from (4.14b) and (4.18).  Thus, as expected
$\overline x_0$ and $\overline y_0$ are the classical coordinates of the
center of the circle in which moves a particle.

We finally point out that the introduced coherent states are stable
with respect to the Hamiltonian $H_\perp$ given by (3.9).  Indeed,
we recall that $x_0$ and $y_0$, and thus $r_{0-}$ are integrals of
the motion.  Further eqs.\ (3.9) and (4.10) yield
\begin{equation}
H_\perp = -\omega \frac{L}{2}.
\end{equation}
Hence, using (4.15) and the first commutator from (4.7) we get
\begin{equation}
Z(t) = e^{\rmi tH_\perp}Ze^{-\rmi tH_\perp}=e^{-\rmi\omega t}Z
\end{equation}
which leads to
\begin{equation}
Z(t) |\zeta,z_0\rangle = \zeta(t) |\zeta,z_0\rangle,
\end{equation}
where $\zeta(t)=e^{-\rmi\omega t}\zeta$.
\section{Comparison with the Malkin-Man'ko coherent states}
In this section we compare the coherent states introduced above and
the Malkin-Man'ko coherent states \cite{1} mentioned in the
introduction using as a test of correctness the closeness to the
classical phase space.  We first briefly sketch the basic properties
of the Malkin-Man'ko coherent states.  Up to an irrelevant
muliplicative constant these states can be defined as the common
eigenvectors of the operators $r_+$ and $r_{0-}$
\numparts
\begin{eqnarray}
&&r_+ |z,z_0\rangle = z |z,z_0\rangle,\\
&&r_{0-}|z,z_0\rangle = z_0 |z,z_0\rangle.
\end{eqnarray}
\endnumparts
Using (4.13c) and (4.13d) we find
\begin{equation}
\langle n,m|z,z_0\rangle = \left(\frac{\mu\omega}{2}\right)^\frac{n}{2}
\frac{z^n}{\sqrt{n!}}\left(\frac{\mu\omega}{2}\right)^\frac{m}{2}
\frac{z_0^m}{\sqrt{m!}}.
\end{equation}
Of course, the states $|z,z_0\rangle$ are the standard coherent
states for the Heisenberg-Weyl algebra generated by the operators $a$,
$a^\dagger$, $b$ and $b^\dagger$ (see (4.9) and (4.11)).  It is also
clear that $z$ and $z_0$ represent the position of a particle on a
circle and the coordinates of the circle center, respectively.  The
parametrization of the complex number $z$ consistent with (4.4) is
of the form
\begin{equation}
z = \overline x + \rmi\overline y,
\end{equation}
where $\overline x$ and $\overline y$ are rectangular coordinates of
a particle on a circle.  Evidently, the parametrization of $z_0$ is
the same as in (4.18).  Now, it follows directly from (5.1a) that
\begin{equation}
\langle r_+\rangle_{(\overline x,\overline y)}=
\frac{\langle \overline x,\overline y;\overline x_0,\overline y_0|r_+|
\overline x,\overline y;\overline x_0,\overline y_0\rangle}
{\langle \overline x,\overline y;\overline x_0,\overline y_0|
\overline x,\overline y;\overline x_0,\overline y_0\rangle}=
\overline x + \rmi\overline y,
\end{equation}
where $|\overline x,\overline y;\overline x_0,\overline y_0\rangle
\equiv|z,z_0\rangle$ with $z$ and $z_0$ given by (5.3) and (4.18),
respectively.  The corresponding formula on the expectation value of
$r_{0-}$ in the normalized coherent state $|\overline x,\overline y;
\overline x_0,\overline y_0\rangle/\sqrt{\langle\overline x,
\overline y;\overline x_0,\overline y_0|\overline x,\overline y;
\overline x_0,\overline y_0\rangle}$ is the same as (4.26).  Using
the polar coordinates we can write (5.4) in the form
\begin{equation}
\langle r_+\rangle_{(l,\varphi)}^{MM}=\langle r_+\rangle_{(\overline
x,\overline y)}=r(l)e^{\rmi\varphi},
\end{equation}
where $r(l)=\sqrt{\overline x^2+\overline y^2}=\sqrt{-\frac{l}
{\mu\omega}}$ following from the classical formula $l=-\mu\omega
r^2~=~-\mu\omega(\overline x^2+\overline y^2)$; the indices MM are
initials for Malkin-Man'ko.  We point out that in opposition to
(4.24) we have the exact relation (5.5).  In this sense the
Malkin-Man'ko coherent states are better approximation of the
configuration space than the states defined by us in the previous
section.  Furthermore, taking into account (4.8) with
$c=-1/(\mu\omega)$, (5.1) and (5.3) we find
\begin{equation}
\langle L\rangle_{(\overline x,\overline y)}=
\frac{\langle \overline x,\overline y;\overline x_0,\overline y_0|L|
\overline x,\overline y;\overline x_0,\overline y_0\rangle}
{\langle \overline x,\overline y;\overline x_0,\overline y_0|
\overline x,\overline y;\overline x_0,\overline y_0\rangle}=
-\mu\omega(\overline x^2+\overline y^2)-1.
\end{equation}
Therefore, using the classical relation $l=-\mu\omega
r^2~=~-\mu\omega(\overline x^2+\overline y^2)$, we get
\begin{equation}
\langle L\rangle_l^{MM}=\langle L\rangle_{(\overline
x,\overline y)}=l-1.
\end{equation}
Thus, it turns out that we have a shift in the classical momentum
and the approximation $\langle L\rangle_l^{MM}\approx l$ is worse in
the light of the observations of section 4 (see discussion under the
formula (4.20)) than the approximate relation $\langle L\rangle_l$
which takes place in the case of the coherent states introduced in
the previous section.  In other words, the coherent states defined
by (4.14) are better approximation of the ``momentum part'' of the
phase space.  We stress that the shift in $l$ in the formula (5.7)
is related to the zero point energy and cannot be ignored.  We
finally remark that as with the states given by (4.14) the
Malkin-Man'ko coherent states are stable with respect to the
evolution generated by the Hamiltonian (3.9).

We now compare the coherent states discussed in section 4 and the
coherent states introduced by Malkin and Man'ko taking as a criterion 
of correctness of the coherent states their closeness to the points
of the classical phase space.  Adopting the idea of the method of
least squares we use as the measure of such closeness the following
entities
\begin{equation}
d(l) = \sqrt{\left(\frac{\langle\!\langle r_+ \rangle\!\rangle_{(l,0)}
-r(l)}{r(l)}\right)^2+\left(\frac{\langle L\rangle_l-l}{l}\right)^2},
\end{equation}
where $\langle\!\langle r_+ \rangle\!\rangle_{(l,\varphi)}$ and
$\langle L\rangle_l$ are given by (4.23) and (4.20), respectively,
for the coherent states defined by (4.14), and analogously
\begin{equation}
d^{MM}(l) = \sqrt{\left(\frac{\langle r_+\rangle_{(l,0)}^{MM}-r(l)}
{r(l)}\right)^2+\left(\frac{\langle L\rangle_l^{MM}-l}{l}\right)^2}
=\frac{1}{|l|},
\end{equation}
for the Malkin-Man'ko coherent states, where in both the above
formulas $r(l)=\sqrt{-\frac{l}{\mu\omega}}$ (see (4.23) and (5.5)).
The distances $d(l)$ and $d^{MM}(l)$ are compared in figure 2.  As
evident from figure 2, the coherent states for a charged particle
in a magnetic field introduced in this paper are better approximations 
of the phase space than the coherent states of Malkin and Man'ko.
\begin{figure*}
\includegraphics[scale=.8]{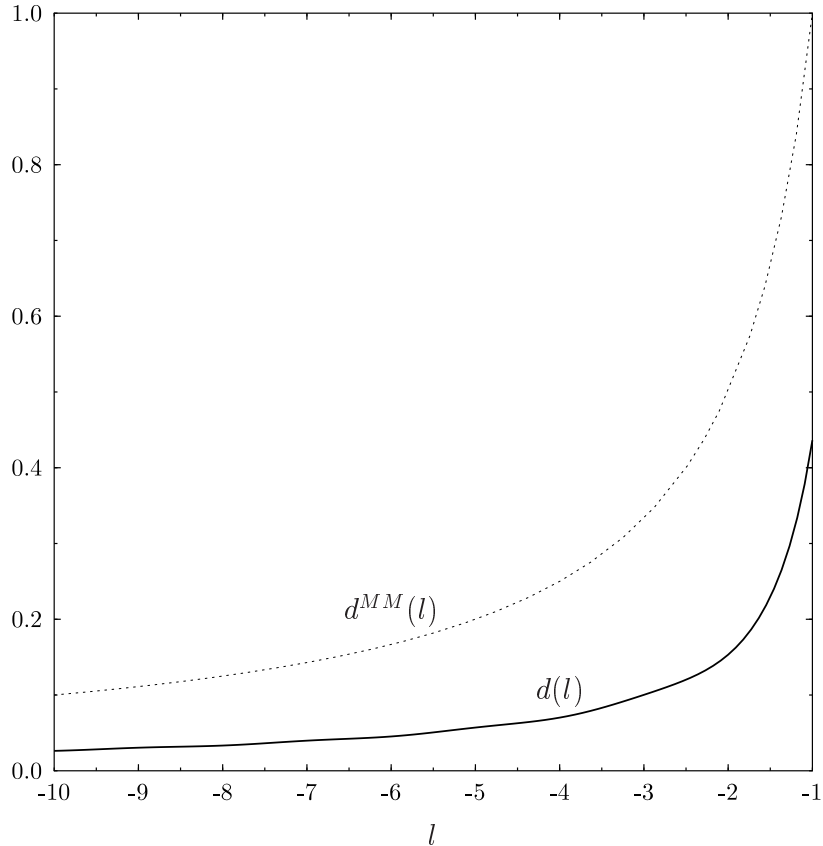}
\caption{Comparison of the closeness to the phase space of the coherent
states introduced in this work (solid line) and the Malkin-Man'ko 
coherent states (dotted line) by means of the distances $d(l)$ and 
$d^{MM}(l)$ given by (5.8) and (5.9), respectively, with $\mu\omega=1$.}
\end{figure*}
\section{Conclusion}
We have introduced in this work the new coherent states for a
charged particle in a uniform magnetic field.  The construction of
these states based on the coherent states for the quantum mechanics
on a circle seems to be more adequate than that of Malkin and
Man'ko.  Indeed, the fact that a classical particle moves transversely
in a uniform magnetic field on a circle, is recognized in the case with
the Malkin-Man'ko coherent states only on the level of the evolution
of these states.  Furthermore, the coherent states introduced in
this work are closer to the points of the classical phase space than
the states discussed by Malkin and Man'ko.  We realize that the best
criterion for such closeness would be minimalization of some
uncertainty relations.  In the case of the coherent states for a
particle on a circle the uncertainty relations have been introduced
by authors in \cite{12} (see also \cite{13} and \cite{14}).  Nevertheless,
the problem of finding the analogous relations for the coherent
states discussed herein seems to be a difficult task.  The reason is
that the radius of the circle is not a c-number as with the coherent
states given by (2.4).  Anyway, in our opinion the simple criterion
of closeness of the coherent states to the points of the classical
phase space based on the definitions (5.8) and (5.9) is precise
enough to decide that the coherent states introduced herein are
better than that discovered by Malkin and Man'ko.  Finally, 
the introduced coherent states should form
a complete set.  We recall that the completeness of coherent states
is connected with the existence, via the ``resolution of the identity
operator'', of the Fock-Bargmann representation.  However, the problem of
finding the resolution of the identity operator is usually nontrivial task.
In our case it is
related to the solution of the problem of moments \cite{15} such that
\begin{displaymath}
\int_{0}^{\infty}x^{n-1}\rho(x)dx = n!e^{(n+\frac{1}{2})^2},
\end{displaymath}
where $\rho(x)$ is unknown density.  Because of the complexity of
the problem the Fock-Bargmann representation for the introduced
coherent states will be discussed in a separate work.
\ack
This paper has been supported by the Polish Ministry of Scientific 
Research and Information Technology under the grant 
No PBZ-MIN-008/P03/2003.

\end{document}